# gScan: Accelerating Graham Scan on the GPU

Gang Mei

Institute of Earth and Environmental Science, University of Freiburg
Albertstr.23B, D-79104, Freiburg im Breisgau, Germany
E-mail: gangmeiphd@gmail.com

**Abstract**
This paper presents a fast implementation of the Graham scan on the GPU. The proposed algorithm is composed of two stages: (1) two rounds of preprocessing performed on the GPU and (2) the finalization of finding the convex hull on the CPU. We first discard the interior points that locate inside a quadrilateral formed by four extreme points, sort the remaining points according to the angles, and then divide them into the left and the right regions. For each region, we perform a second round of filtering using the proposed preprocessing approach to discard the interior points in further. We finally obtain the expected convex hull by calculating the convex hull of the remaining points on the CPU. We directly employ the parallel sorting, reduction, and partitioning provided by the library *Thrust* for better efficiency and simplicity. Experimental results show that our implementation achieves 6x ~ 7x speedups over the Qhull implementation for 20M points.

## 1. Introduction

Given a set of planar points S, the 2D convex hull problem is to find the smallest polygon that contains all the points in S. This problem is a fundamental issue in computer science and has been studied extensively. Several classic convex hull algorithms have been proposed [1-7]; most of them run in O(nlogn) time. Among them, the Graham scan algorithm [1] is the first practical convex hull algorithm, while QuickHull [7] is the most efficient and popular one in practice.

Recently, the computational capability of the GPU has surpassed that of the CPU and is being used to solve large-scale problems in various applications. There are also some attempts at solving the convex hull problem on the GPU. For example, Srikanth, et al. [8] used NVIDIA GPU and Cell BE hardware to accelerate the construction of 2D convex hulls by adopting the well-known QuickHull approach [7]. Similarly, Srungarapu, et al. [9] and Jurkiewicz and Danilewski [10] parallelized the QuickHull algorithm to accelerate the finding of two dimensional convex hulls.

Also by adopting the QuickHull, Stein, et al. [11] presented a novel parallel algorithm for computing the convex hull of a set of points in 3D. Tang, et al. [12] developed a CPU-GPU hybrid algorithm to compute the convex hull of points in three or higher dimensional spaces. Tzeng and Owens [13] presented a framework for accelerating the computing of convex hull in the divide-and-conquer fashion by taking advantage of QuickHull. Similarly, White and Wortman [14] described a pure GPU divide-and-conquer parallel algorithm for computing 3D convex hulls based on the Chan's minimalist 3D convex hull algorithm [15]. By exploiting the relationship between Voronoi diagram and convex hull, Gao, et al. [16] introduced a two-phase 3D convex hull algorithm. In addition, Gao, et al. [17] designed ffHull, a flip algorithm that allows nonrestrictive insertion of many vertices before any flipping of edges.

In this paper, we present a fast implementation of the Graham scan algorithm on the GPU. There are typically two stages in the Graham scan algorithm: the first stage is to sort points according to their angles and the second is to loop over all the sorted points in sequence to determine extreme points. The efficiency bottleneck of the Graham scan is the sorting of points. An effective strategy for improving computational efficiency is to discard the interior points that need to be sorted. The simplest case in two dimensions is to form a convex quadrilateral using four extreme points with min or max *x* or *y* coordinates and then check each point to determine whether it locates inside the quadrilateral; see [4]. Another quite recent effort for efficiently discarding interior points was introduced in [18].

We employ the above strategy to improve the efficiency of our algorithm. In addition to the simple preprocessing procedure introduced in [4], we also propose a novel preprocessing approach that is well suitable to be used in the Graham scan. In our algorithm, we perform the following two efforts: (1) we apply effective preprocessing procedures by discarding interior points to reduce the number of points that need to be sorted, and (2) we sort the remaining points extremely fast using the efficient parallel sorting algorithm.

Similar to the traditional Graham scan, our GPU-accelerated convex hull algorithm is also composed of two stages. The first stage mainly includes two rounds of preprocessing procedures and a fast sorting of points, which is performed on the GPU. The second stage is the finalization of finding the convex hull, which is carried out on the CPU. Our algorithm is implemented by heavily taking advantage of the library *Thrust* [19] for better efficiency and simplicity.

The rest of this paper is organized as follows. Section 2 introduces the outline and several basic ideas of our 2D convex hull algorithm, while Section 3 details our implementation techniques. The experimental results are presented in Section 4 and we discuss the results in Section 5. Finally, Section 6 concludes our work.

## 2. Our Algorithm

### 2.1 Algorithm Design

The proposed GPU-accelerated convex hull algorithm is designed on the basis of the Graham scan. There are typically two stages in the Graham scan algorithm: the first stage is to sort points according to their angles and the second is to loop over all the sorted points in sequence to determine extreme points. The efficiency bottleneck of the Graham scan is the sorting of points according to their angles.

When computing the convex hull of a large set of points, the sorting of all points is quite time-consuming and thus the overall procedure of finding the convex hull could be very slow.

To handle the above problem and improve the efficiency, we perform the following two efforts on the GPU to speed up the calculating in the first stage: (1) we apply effective preprocessing procedures by discarding interior points to reduce the number of points that need to be sorted, and (2) we sort the remaining points extremely fast using the efficient GPU-based parallel sorting algorithm.

Similar to the traditional Graham scan, our algorithm is also composed of two stages. The first stage mainly includes two rounds of preprocessing procedures and a fast sorting of points, which is performed on the GPU. The second stage is the finalization of finding the convex hull, which is carried out on the CPU. More specifically, the procedure of the proposed algorithm is listed as follows:

1) Find four extreme points that have the max or min $x$ or $y$ coordinates by parallel reduction, denote them as $P_{minx}$, $P_{maxx}$, $P_{miny}$, $P_{maxy}$
2) Determine the distribution of all points in parallel, and discard the points locating inside the quadrilateral formed by $P_{minx}$, $P_{miny}$, $P_{maxx}$, and $P_{maxy}$
3) Calculate the distance and angle of each point
4) Sort all points in the ascending order of angles
5) Find the point that has the longest distance, denote it as $P_{longest}$
6) Divide the list of sorted points into the left and the right region using the point $P_{longest}$; see Figure 1(a).
7) Perform the proposed preprocessing approach for both the right and the left regions to discard the interior points in further
8) Find the convex hull of the remaining points

The first round of preprocessing procedure by discarding the interior points locating inside a quadrilateral is first carried out (i.e., the Step 1 and Step 2); then the sorting of the remaining points is performed after calculating the distance and angle of each point (Step 3 and Step 4). In the Steps 5 ~ 7, the second round of preprocessing is employed to in further discard interior points for the pre-sorted points. These 7 steps are completely performed on the GPU, while the final step, i.e., the Step 8, is carried out on the CPU to finalize the finding of the convex hull.

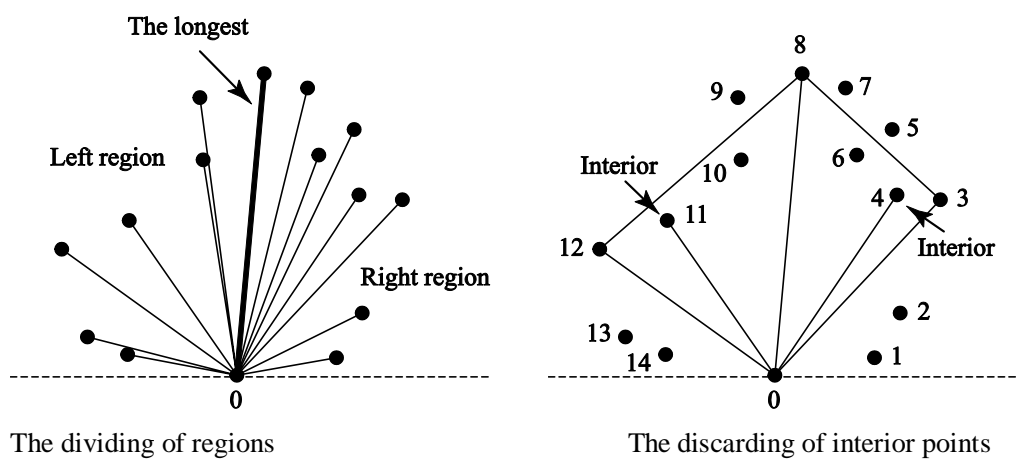

The dividing of regions    The discarding of interior points

Figure 1 The preprocessing procedure for checking and discarding interior points

## 2.2 The Second Round of Preprocessing (Discarding)

2.2.1 Overview of the Proposed Preprocessing Approach

This section presents a novel preprocessing procedure for discarding the interior points. The basic idea behind this approach is to take advantage of the order of sorted points for checking whether a point locate inside a triangle. Typically, when checking whether a point locate in a triangle, it needs to determine whether the point is on the left side of each segment of the triangle. For example, in Figure 1 (b), there are totally 15 points; and 14 points have been sorted according to their angles. Taking the point P4 for an example, it generally needs to check whether P4 is on the left side of the directed line P0P3, P3P8, and P8P0. However, due to points have been sorted in the ascending order of angles, the angle of P4 is larger than that of the point P3 and less than that of the point P8. Therefore, it is obviously that P4 is on the left side of the directed line P0P3, and on the right side of the directed line P0P8. Hence, in this case it is only to check whether P4 is on the left side of the directed line P3P8 for determining whether P4 locates in the triangle P0P4P8.

As a summary, it only needs to check the point P4 whether it locates on the left side of the directed line P3P8 for determining whether it is an interior point. Similarly, for the point P11, it only needs to check whether it locates on the right side of the directed line P12P8. By taking advantage of the order of the sorted points, the overhead of computations can be reduced. The pseudo-code of this approach is listed in Figure 2. Two simple examples are presented in Figure 3 and Figure 4 for demonstrating the discarding of interior points for the right region and the left region, respectively.

```
n = number of points
l = index of the point with the longest distance
P_temp ← P_0
for i = 1 to i < l do
    if P_i is on the left side of the line P_temp P_l,
    then point P_i is interior
    else P_temp ← P_i
end

P_temp ← P_{n-1}
for i = n - 2 to i > l do
    if P_i is on the right side of the line P_temp P_l,
    then point P_i is interior
    else P_temp ← P_i
end
```

Figure 2 The procedure of the proposed preprocessing approach

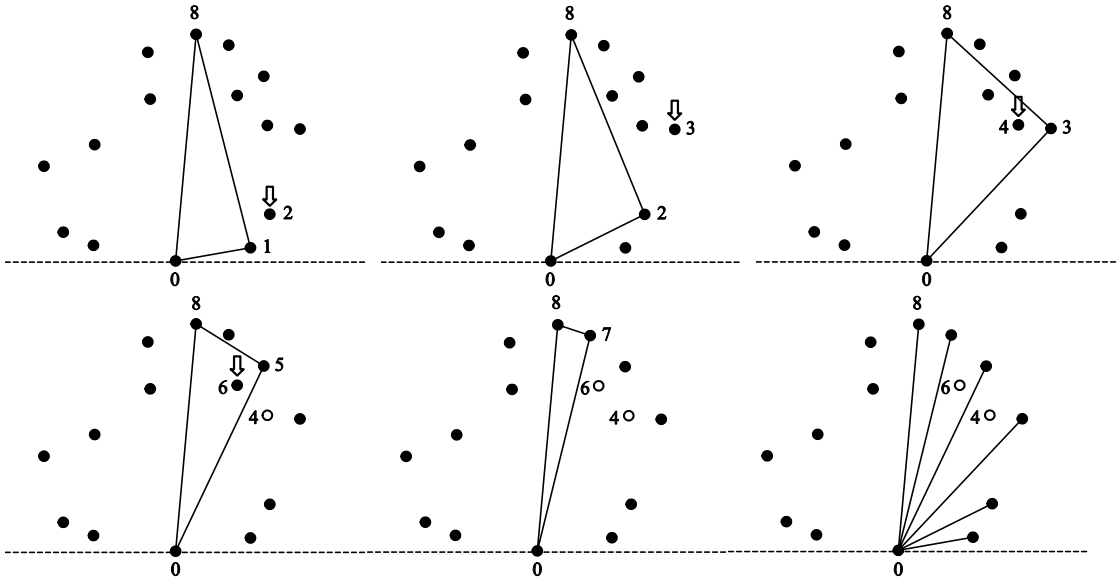

Figure 3 A simple example of discarding interior points in the right region

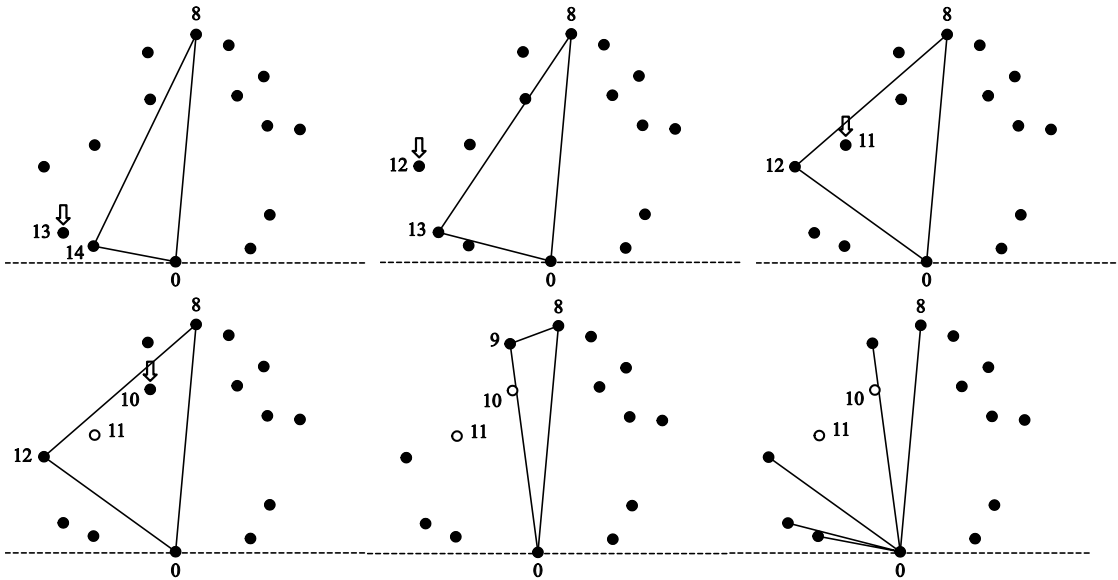

Figure 4 A simple example of discarding interior points in the left region

### 2.2.2 Correctness of the Preprocessing Approach

It is clear that: in the finding of the convex hull for a set of planar points, if a point locate inside a convex polygon such as a triangle or a quadrilateral formed by other points, then this point must be an interior point, and can be discarded. In Section 2.2.1, we have explained that it only needs to check whether a point locate on the left or right side of a directed line for determining whether it is an interior point. In the proposed preprocessing approach (demonstrated in Figure 3 and Figure 4), we only discard those points that have been exactly determined to be interior ones. We do not discard any potential extreme points. Hence, the correctness of our implementation can be guaranteed.

### 2.2.3 Data Dependency and Divide-and-Conquer

According to the proposed preprocessing approach (Figure 2), data dependencies exist in the discarding of interior points for that the point $P_{temp}$ needs to be dynamically determined. For example, after checking the point P2, the point $P_{temp}$ is assigned to P2 for checking the point P3. When checking the point P4, the point $P_{temp}$ is exactly the point P3; however, after checking the point P4, the point $P_{temp}$ is still the point P3. The point $P_{temp}$ cannot be independently determined when checking each point.

Due to the data dependencies in the proposed preprocessing approach, it is well suited to be implemented in the sequential programming pattern. To implement this approach in parallel, an effective solution is to first divide the large set of pre-sorted points into some smaller subsets of points, and then check the interior points for all subsets of points in parallel. For an individual subset of points, the checking of interior points is still performed in sequence. This solution is in the *Divide-and-Conquer* fashion.

## 3. Implementation Details

There are two main stages in our implementation: the first stage mainly includes two rounds of preprocessing procedures by discarding interior points on the GPU; and the second stage is the finalization of finding the convex hull of the remaining pre-sorted points on the CPU. The implementing of the second stage is much easier and simpler than that of the first stage; thus, in this section, implementation details are focused on the first stage.

**3.1 The First Round of Discarding**

The basic idea behind this round of discarding is quite simple. Four extreme points with min or max x or y coordinates can be easily found and then used to form a quadrilateral; any points that locate inside the quadrilateral can be directly discarded. This round of discarding is easily implemented in the sequential programming pattern.

When implementing this round of discarding on the GPU, there are obviously two sub procedures that can be well mapped and performed in the parallel programming pattern on the GPU: the first procedure is the finding of the points with min or max x or y coordinates; and the second is the checking whether each point locates inside the quadrilateral.

The finding of the min or max values can be quite efficiently realized using the parallel reduction on the GPU. *Thrust* provides such efficient parallel primitive. In our implementation, we use the interface function thrust::min_element() and thrust::max_element() to find the points with min or max coordinates.

The checking of each point to determine whether it falls in the quadrilateral does not have any data dependencies. In other words, the checking for all points can be well performed in parallel. We design a simple CUDA kernel to parallelize this checking. Each thread is responsible for determining whether a point locate inside the quadrilateral. An array int pos[n] is allocated on the GPU to store the indicator values of all points. If a point is in the quadrilateral, then its indicator value will be set to 0; otherwise, 1.

After determining the distribution of all points, all the interior points should be removed from the input list of points. This removal can also be efficiently carried out in parallel. We first use the function thrust::partition () to gather all interior points together, then remove the interior points and keep the remaining points. Those remaining points will be used to calculate the convex hull in subsequent steps.

### 3.2 The Calculating of Distances and Angles

The calculating of the distance and angle for each point is straightforward. We also design a kernel to compute the distances and angles in parallel. Each thread within the thread grid is invoked for calculating the distance and angle for each point. Results are stored in two arrays, dist[n] and angle[n]. The angles will be used to sort points and the distances will be employed to perform the second round of discarding.

### 3.3 The Sorting of Points

To speed up the sorting of the remaining points according to the angles, we also use the efficient function provided by thrust, i.e., thrust::sort_by_keys (). The keys that are used for sorting is the angles of points. Several zip_iterators are created to combine the coordinates, angle, and distance of all points into a virtual array of structures. This sorting based on 32-byte keys is extremely fast and thus can improve the efficiency of the entire implementation.

### 3.4 The Second Round of Discarding

We design a kernel for each region to discard the interior points using the proposed preprocessing approach; more details about the preprocessing approach are introduced in Section 2.2. There is only one thread block within the kernel's thread grid. Each thread in the only thread block is responsible for checking consecutive ($m$ + BLOCK_SIZE - 1) / BLOCK_SIZE points in the same region, where $m$ is the number of points in a region for being checked, and BLOCK_SIZE represents the number of threads in the only block. In our implementation, we set BLOCK_SZIE to 1024 according to the compute capability of the adopted GPU. After checking and discarding interior points in this round, some previous exterior points have been determined as interior ones; and their corresponding indicator values are modified to 0.

In our implementation, we only allocate one thread block in the discarding of interior points. This is due to the data dependency issues in the discarding. More details and demonstrations about the data dependency are introduced in Section 2.2. As described in Section 2.2.3, the checking for a set of consecutive points can only be performed in a sequential pattern. However, it is able to first divide a large set of consecutive points into some smaller subsets of consecutive points, and then perform the checking in parallel for each subset of points separately. We adopt this solution; but we cannot determine the optimal size of a subset of points or the number of all subsets. Thus, we decide to divide a large set of consecutive points into BLOCK_SIZE subsets, while each subset contains ($m$ + BLOCK_SIZE - 1) /

BLOCK_SIZE points; and then, we carry out the checking for all the BLOCK_SIZE subsets in parallel.

After checking and updating the indicator value for each point, we perform another parallel reduction to remove the interior points that have been determined in this round of discarding. Noticeably, the remaining points have been sorted in the ascending order of their angles. These pre-sorted points can be directly used to calculate the convex hull using the method introduced in the traditional Graham scan. To preserve the relative order of the sorted points, we use the function `thrust::stable_partition()` rather than `thrust::partition()` to divide the interior points and the exterior points into separate ranges, and then use the function `thrust::copy()` to copy the coordinates of the exterior points for outputting.

## 4. Results

We have tested our implementation against the Qhull library [20] using three groups of test data on two different platforms. More details about the platforms are listed in Table 1. We have created three groups of datasets for testing. The first group includes 5 sets of randomly distributed points in a square that are generated using the `rbox` component in Qhull. Similarly, the second group is composed of 5 sets of randomly distributed points in a circle. The third group consists of 5 point sets that are derived from 3D mesh models by projecting the vertices of each 3D model onto the XY plane. These mesh models presented in Figure 5 are directly obtained from the Stanford 3D Scanning Repository[1] and the GIT Large Geometry Models Archive[2].

**Reuse of Some Experimental Results**: the running time of the baseline implementation, Qhull, for the first group of test data (see Table 2 and Table 3) and the third group of test data (see Table 6 and Table 7) have been presented in [21]. We directly adopt this part of the results. Any other results are original.

Table 1 The platforms used for testing

|  | Platform no.1 | Platform no.2 |
| --- | --- | --- |
| CPU | Intel i7-3610QM (2.30GHz) | Intel i5-3470 (3.20GHz) |
| Memory | 6GB | 8GB |
| GPU | GeForce GTX 660M | GeForce GT640 (GDDR5) |
| GPU RAM | 2GB | 1GB |
| GPU Cores | 384 | 384 |
| GPU Compute Capability | 3.0 | 3.5 |
| CUDA | Version 6.0 | Version 6.0 |
| OS | Window 7 | Window 7 |

---

[1] http://www-graphics.stanford.edu/data/3Dscanrep/

[2] http://www.cc.gatech.edu/projects/large_models/

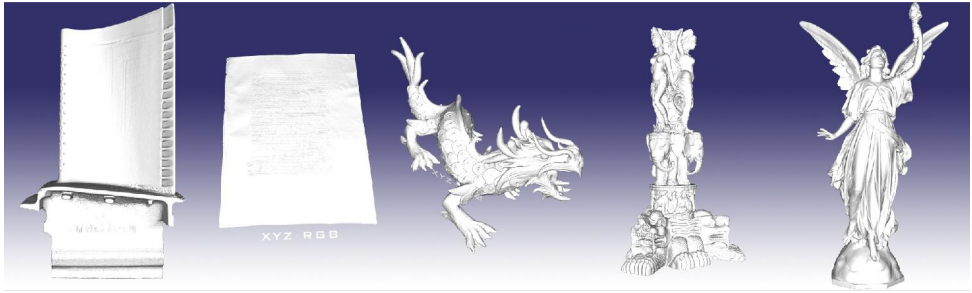

**Figure 5** 3D mesh models from Stanford 3D Scanning Repository and GIT Large Geometry Models Archive. From the left to the right, the models are: Turbine Blade, Vellum Manuscript, Asian Dragon, Thai Statue, and Lucy. (This picture is also presented in [21])

### 4.1 Efficiency of the gScan

**4.1.1 The First Group of Tests: Points in a Square**

For the first groups of test data, i.e., the group of randomly distributed point sets in a square, the running time on the GPU GTX 660M and GT 640 is listed in Table 2 and Table 3, respectively. Comparisons of efficiency are presented in Figure 6. To evaluate the computation load between the GPU side and the CPU side of our algorithm, we count the running time separately for both of the two sides and calculate the workload percentage of the CPU side.

Experimental results show that: on both of the GPUs, the speedups of gScan over Qhull become larger with the increasing of the data size; however, the increasing of the speedup is not significant. And the speedup is about 3x~4x on average. The workload percentage of the CPU side is much smaller than that on the GPU side; and it sustainably decreases when the data size increases. In addition, the workload percentage of the CPU side is usually less that 10%.

**Table 2** Comparison of running time (/ms) for the point sets locating in squares on GTX 660M

| Size | Qhull | gScan | | | | Speedup |
| --- | --- | --- | --- | --- | --- | --- |
| | | Total | GPU | CPU | CPU(%) | |
| 1M | 237 | 60.9 | 56.1 | 4.8 | 7.88 | 3.89 |
| 2M | 426 | 103.6 | 97.2 | 6.4 | 6.18 | 4.11 |
| 5M | 605 | 134.1 | 127.3 | 6.8 | 5.07 | 4.51 |
| 10M | 1171 | 283.3 | 276.4 | 6.9 | 2.44 | 4.13 |
| 20M | 2353 | 482.0 | 472.7 | 9.3 | 1.93 | 4.88 |

**Table 3** Comparison of running time (/ms) for the point sets locating in squares on GT 640

| Size | Qhull | gScan | | | | Speedup |
| --- | --- | --- | --- | --- | --- | --- |
| | | Total | GPU | CPU | CPU(%) | |
| 1M | 109 | 33.2 | 31.4 | 1.8 | 5.42 | 3.28 |
| 2M | 202 | 60.7 | 57.9 | 2.8 | 4.61 | 3.33 |
| 5M | 515 | 116.9 | 114.1 | 2.8 | 2.40 | 4.41 |
| 10M | 1034 | 249.5 | 246.6 | 2.9 | 1.16 | 4.14 |
| 20M | 2215 | 446.4 | 442.9 | 3.5 | 0.78 | 4.96 |

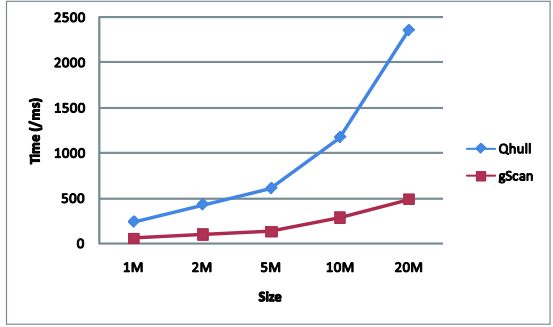
GTX 660M

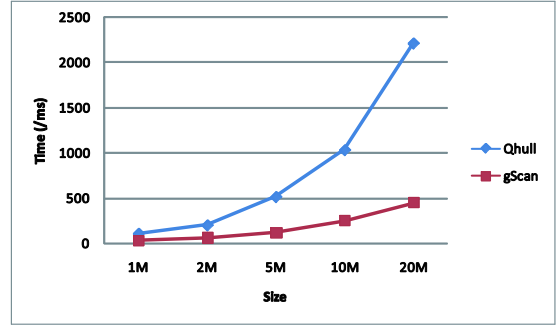
GT 640

Figure 6 The efficiency of gScan against CPU-based Qhull for the point sets distributed in squares on two platforms

### 4.1.2 The Second Group of Tests: Points in a Circle

The running time of the second group of test data on two GPUs is presented in Table 4 and Table 5, and compared in Figure 7. In this group of test data, each set of points is randomly distributed in a circle rather than in a square. The testing results show that the speedup of gScan over Qhull on the GPU GTX 660M is a bit larger than that on the GPU GT 640. The speedup is about 5x~6x on average and 6x~7x in the best cases.

These results also indicate that our implementation gScan can achieve better efficiency for those sets of points in circles (i.e., the second group of test data) than that for the sets of points in squares (i.e., the first group of test data). In addition, the workload percentage of the CPU side for the second group of test data less is always less than that for the first group of test data.

Table 4 Comparison of running time (/ms) for the point sets locating in circles on GTX 660M

| Size | Qhull | gScan | | | | Speedup |
|---|---|---|---|---|---|---|
| | | Total | GPU | CPU | CPU(%) | |
| 1M | 225 | 54.6 | 52.8 | 1.8 | 3.30 | 4.12 |
| 2M | 430 | 68.4 | 66.4 | 2.0 | 2.92 | 6.29 |
| 5M | 982 | 140.7 | 137.9 | 2.8 | 1.99 | 6.98 |
| 10M | 1897 | 253.1 | 249.1 | 4.0 | 1.58 | 7.50 |
| 20M | 3811 | 482.4 | 475.8 | 6.6 | 1.37 | 7.90 |

Table 5 Comparison of running time (/ms) for the point sets locating in circles on GT 640

| Size | Qhull | gScan | | | | Speedup |
|---|---|---|---|---|---|---|
| | | Total | GPU | CPU | CPU(%) | |
| 1M | 134 | 34.4 | 33.3 | 1.1 | 3.20 | 3.90 |
| 2M | 258 | 55.3 | 54.1 | 1.2 | 2.17 | 4.67 |
| 5M | 652 | 116.2 | 114.9 | 1.3 | 1.12 | 5.61 |
| 10M | 1337 | 216.2 | 214.7 | 1.5 | 0.69 | 6.18 |
| 20M | 2626 | 418.1 | 416.3 | 1.8 | 0.43 | 6.28 |

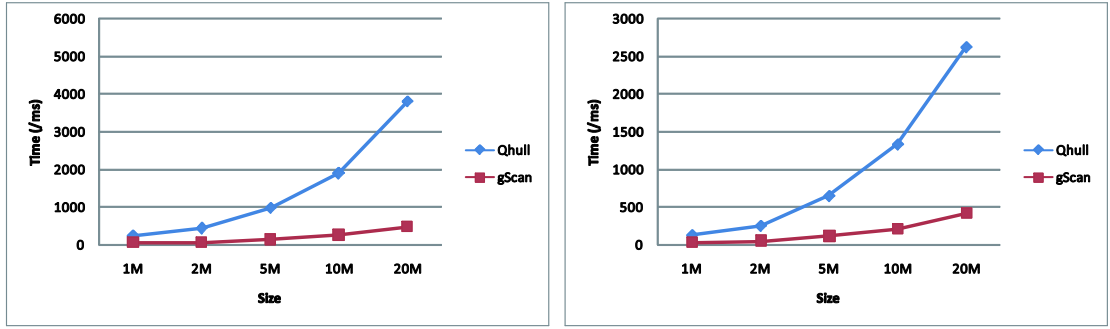

GTX 660M                                                                 GT 640

Figure 7 The efficiency of gScan against CPU-based Qhull for the point sets distributed in circles on two platforms

**4.1.3 The Third Group of Tests: Points Derived from Mesh Models**

This group of test data consists of 5 point sets that are derived from 3D mesh models by projecting the vertices of each 3D model onto the XY plane. The experimental results of this group are presented in Table 6, Table 7, and Figure 8.

On both of the two GPUs, the speedup of gScan over Qhull is about 5x~6x on average and 6x~7x in the best cases, which is larger than that for the first group of test data but less than that for the second group of test data. Comparing the results on two GPUs, the speedups on the GPU GTX660M is always larger than those on the GPU GT 640. The workload percentage of the CPU side also sustainably decreases when the data size increases.

**Table 6** Comparison of running time (/ms) for the point sets derived from models on GTX 660M

| Model | Size | Qhull | gScan | | | | Speedup |
|---|---|---|---|---|---|---|---|
| | | | Total | GPU | CPU | CPU(%) | |
| Blade | 0.8M | 202 | 52.5 | 47.4 | 5.1 | 9.71 | 3.85 |
| Vellum | 2.1M | 392 | 73.0 | 69.9 | 3.1 | 4.25 | 5.37 |
| Asian Dragon | 3.6M | 492 | 85.8 | 82.5 | 3.3 | 3.85 | 5.73 |
| Thai Statue | 5M | 547 | 91.2 | 89.1 | 2.1 | 2.30 | 6.00 |
| Lucy | 14M | 1481 | 213.8 | 210.2 | 3.6 | 1.68 | 6.93 |

**Table 7** Comparison of running time (/ms) for the point sets derived from models on GT 640

| Model | Size | Qhull | gScan | | | | Speedup |
|---|---|---|---|---|---|---|---|
| | | | Total | GPU | CPU | CPU(%) | |
| Blade | 0.8M | 78 | 31.1 | 28.8 | 2.3 | 7.40 | 2.51 |
| Vellum | 2.1M | 218 | 62.1 | 59.8 | 2.3 | 3.70 | 3.51 |
| Asian Dragon | 3.6M | 343 | 79.8 | 77.5 | 2.3 | 2.88 | 4.30 |
| Thai Statue | 5M | 468 | 82.1 | 80.0 | 2.1 | 2.56 | 5.70 |
| Lucy | 14M | 1295 | 207.6 | 205.2 | 2.4 | 1.16 | 6.24 |

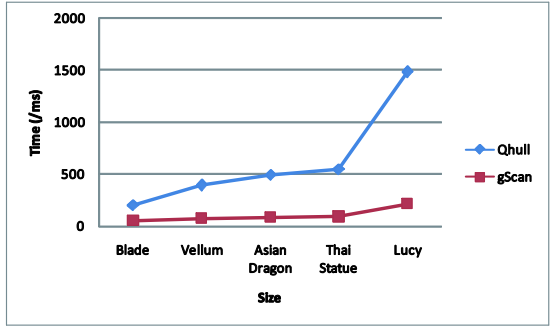 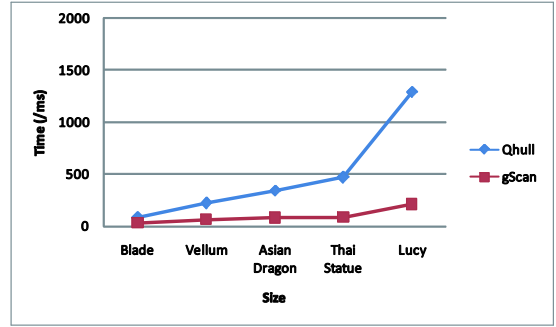

GTX 660M                                                          GT 640

Figure 8 The efficiency of gScan against CPU-based Qhull for the point sets derived from 3D mesh models on two platforms

## 4.2 Effectiveness of the Second Round of Discarding

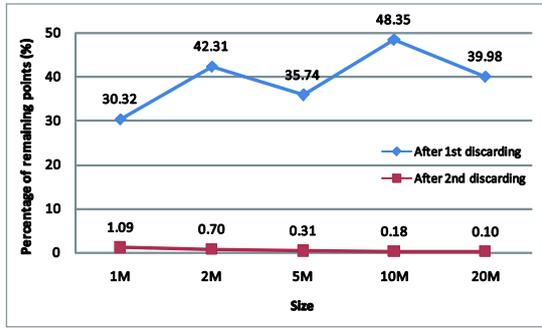 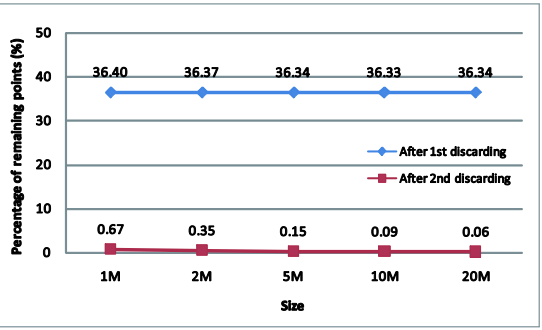

(a) Randomly distributed points in squares    (b) Randomly distributed points in circles

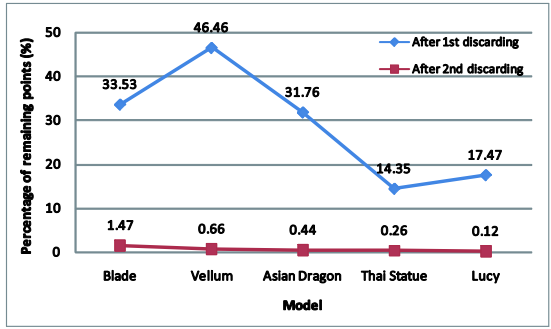

 (c) Points derived from 3D models

Figure 9 The effectiveness of two rounds of discarding interior points (compared by percentages)

There are two rounds of discarding in our algorithm. The first round of discarding is based on four extreme points with min or max x or y coordinates. We propose an angle-based preprocessing approach for the second round of discarding. To evaluate the effectiveness of the proposed preprocessing approach, we count the remaining points after each round of discarding, calculate the percentages, and compare the effectiveness of two rounds of discarding. The results presented in Figure 9 show that our preprocessing approach can dramatically reduce the number of remaining points and thus improve the overall efficiency of gScan. In addition, the effectiveness of the second round of discarding becomes better with the increasing of the data size. Comparing the results generated for those three groups of test data, the results for the group of sets of points in circles are the best; see Figure 9(b).

# 5 Discussion

## 5.1 Comparison

We have compared our implementation gScan with the library Qhull [20] using three groups of test data. Experimental results show that our implementation outperforms the Qhull, and can achieve 5x~6x speedups on average and 6x~7x speedups in the best cases over the library Qhull. The efficiency gains benefit from: (1) the parallelization of several procedures on the GPU and (2) the reduction of points being used to calculate the convex hull by discarding interior points.

Compared to other GPU-accelerated convex hull algorithm such as those were designed on the basis of QuickHull [8, 9, 11, 13], the main advantage of our algorithm is that it is very easy to implement, which is mainly due to (1) the use of the library Thrust and (2) relatively less data dependencies. The data-parallel primitives such as parallel sorting and parallel reduction provided by Thrust are very efficient and easy to use; we can directly use these primitives in CUDA to realize our implementation without too many efforts. In addition, in our algorithm the only step that has data dependency is the second round of checking and discarding interior points. Other steps or procedures can be very well mapped to the massively parallel nature of the modern GPU. This feature of having less data dependencies also makes our algorithm simple and easy to implement in practical applications.

Compared to a similar algorithm introduced in [21], there are several the same ideas, including: (1) the performing of two rounds of preprocessing procedures by discarding interior points on the GPU, and (2) the finalization of computing the convex hull on the CPU. More specifically, the algorithm presented in [21] and the proposed algorithm in this paper employ the idea of discarding interior points by checking whether each point locate inside a quadrilateral formed by several (typically four) extreme points. In addition, both of the two algorithms need to sort the remaining points before carrying out the second round of discarding.

However, our algorithm presented in this paper is different from the algorithm introduced in [21]. The main difference is the underlying algorithm that is employed for parallelizing. The proposed algorithm in this paper, gScan, is designed on the basis of the Graham scan [1], while the algorithm in [21], i.e., CudaChain, is developed based on the fast convex hull algorithm proposed by Akl and Toussaint [4].

The second difference is the sorting of points before carrying out the second round of discarding. Although in both algorithms the remaining points are needed to be sorted before performing the second round of discarding, those points are sorted according to their angles in gScan, while in the algorithm CudaChain points are sorted according to their coordinates. In addition, it needs to sort the remaining points only once in gScan; in contrast, the remaining points are needed to be sorted four times in CudaChain.

The third difference is the criteria for determining which points are interior ones in the second round of discarding. In gScan, the remaining points are first sorted and then divided into the left and the right parts. For both of the two parts, each point is checked according to its location with respect to a direct line to determine whether it is an interior point. In CudaChain, the remaining points are first distributed into four sub regions (i.e., the lower left, the lower right, the upper right, and the upper left), and then sorted separately. For each sub region, the x or y coordinate of a point is used to compare with those of other points to determine whether it is an interior point.

We summary the similarities and differences between the gScan and CudaChain; see Table 8.

Table 8 Comparison of CudaChain with gScan

| Algorithm Design and Details | CudaChain [21] | gScan |
|---|---|---|
| Stages | First stage on the GPU<br>Second stage on the CPU | The same |
| The first round of discarding | Based on 4 extreme points | The same |
| The second round of discarding | Need to previously sort point | The same |
| Underlying algorithm | Akl and Toussaint [4] | Graham scan [1] |
| Sorting of points | According to x or y | According to angles |
| Criteria in 2nd discarding | Compare coordinates | Check whether in a triangle |
| Sorting and distribution | Distribute points into sub regions first and then sort | Sort points first, then divide points into two sub regions |
| Number of sub regions | 4 | 2 |
| Number of sorting | 4 | 1 |
| Number of partitioning | 5 | 2 |

## 5.2 Correctness and Complexity

The correctness of our algorithm can be guaranteed obviously. As mentioned several times, there are two main stages in our implementation. If both of the two stages can be proved to be correct, then the entire implementation is definitely correct. In the first stage, we perform two rounds of preprocessing procedures by discarding interior points. The first round of preprocessing is to discard those points that locate inside a quadrilateral formed by typically four extreme points. This round of discarding is obviously correct and does not need any proofs. For the second round of discarding, i.e., the proposed preprocessing approach based on pre-sorted angles, its correctness has been proved in Section 2.2.2; see Figure 1. Thus, the first stage of our algorithm can be guaranteed to be correct. The second stage is performed on the CPU to calculate the convex hull of the sorted points, which is the same as the finding of the convex hull for the radially sorted points in the traditional Graham scan. Thus, this stage can also be guaranteed to be correct.

In summary, our algorithm can be guaranteed to be correct since: (1) we only discard those points that have been correctly determined to be interior ones and do not remove any potential extreme points; (2) we use all the remaining points to finalize the finding of the convex hull.

The time complexity of our algorithm in the worst case is O(nlogn). The O(nlogn) is spent sorting the points according to the angles. Both the first and the second rounds of discarding run in O(n) time. In addition, the second stage of our algorithm employs a stack-based method which also runs in O(n) time. Therefore, the time complexity of the entire algorithm is O(nlogn).

**5.3 Limitation and Future Work**

An obvious shortcoming of our algorithm is that it cannot be applied in the case where all input points are extreme points. For example, all the input points locate on a circle. One of the basic ideas behind our algorithm is to improve the efficiency by discarding interior points. We perform two rounds of discarding in the first stage of our algorithm. When all input points are extreme ones, no interior points can be found and all the input points are used to calculate the convex hull. In this case, the two rounds of discarding are thus wasteful; and the entire algorithm is inefficient.

When developing our implementation gScan, we have used several efficient data-parallel algorithm primitives provided by Thrust such as sorting, reduction, and partitioning. The efficiency of our implementation is heavily depends on the efficiencies of such parallel primitives. Recently, a similar but more efficient counterpart of the library Thrust, CUB [22], has been developed. We expect to gain a significant increase in efficiency performance of our implementation by replacing the primitives provided by Thrust with those corresponding primitives in CUB. Future work should therefore include the implementation our algorithm using CUB and the evaluation of efficiency performance.

# 6. Conclusion

We have presented a GPU-accelerated implementation of the Graham scan algorithm in this paper. We have also proposed an effective preprocessing procedure for discarding interior points. Our implementation is mainly composed of two stages: the first stage that includes two rounds of preprocessing procedures on the GPU and the second stage where the calculating of the convex hull for the remaining points is finalized on the CPU. Our implementation is developed by exploiting the library Thrust. Several efficient data-parallel algorithm primitives such as parallel sorting, reduction, and partitioning provide by Thrust are used to make our implementation simple to implement and easy to use. We have compared our implementation to the library Qhull using three groups of test data on two different platforms. Experimental results show that our implementation can achieve the speedups of 5x ~ 6x on average and 6x ~ 7x in the best cases over the library Qhull. We believe that our implementation is competitive in practical applications for its simplicity and satisfied efficiency.

## Acknowledgements
We thank the academic editor and the reviewers for their helpful comments and suggestions. The first author Gang Mei would like to acknowledge the PhD scholarship provided by the China

Scholarship Council (CSC) to support his studying in Germany.

**Conflict of Interests**: The author declares that there is no conflict of interests regarding the publication of this article.